\documentclass[12pt,showpacs,preprintnumbers,amsmath,amssymb,nofootinbib,superscriptaddress]{article}
\usepackage{amssymb}
\usepackage{amscd}

\begin{document}
\begin{titlepage}

{\hbox to\hsize{\hfill December 2010 }}

\bigskip \vspace{3\baselineskip}

\begin{center}
{\bf \large 
Cosmological constant in scale-invariant theories}

\bigskip

\bigskip

{\bf Robert Foot, Archil Kobakhidze and Raymond R. Volkas \\}

\smallskip

{ \small \it
School of Physics, The University of Melbourne, Victoria 3010, Australia \\
E-mails: rfoot@unimelb.edu.au, archilk@unimelb.edu.au, raymondv@unimelb.edu.au
\\}

\bigskip
 
\bigskip

\bigskip

{\large \bf Abstract}

\end{center}
\noindent 
The incorporation of a small cosmological constant within radiatively-broken 
scale-invariant models is discussed.  We show that phenomenologically consistent scale-invariant models can be constructed which allow a small positive cosmological constant, providing certain relation between the particle masses is satisfied. As a result, the mass of the dilaton is generated at two-loop level. 
Another interesting consequence is that the electroweak symmetry-breaking vacuum in such models is necessarily a metastable `false' vacuum which, fortunately, is not expected to decay on cosmological time scales.

\end{titlepage}

\section{Introduction}

In recent years, we have proposed a number of simple perturbative 
extensions of the standard model \cite{Foot:2007as}-\cite{Foot:2010av} where mass 
scales are generated radiatively \cite{Coleman:1973jx} through the phenomenon of dimensional transmutation. 
This phenomenon generates a certain fundamental mass scale $\Lambda$, even though the classical
version of the theory is exactly scale invariant (this is the scale anomaly).  The scale $\Lambda$ is defined
through a renormalization condition 
that we shall explain in the next section.  This quantally-induced scale is a fundamental parameter in the
quantized version of the theory, and it replaces one of the dimensionless parameters of the classical theory.
But we know that many different mass scales are needed to describe nature, including the electroweak,
see-saw and Planck scales.  How can different scales arise in a framework where there is only one fundamental
mass parameter?  One answer is that the various scales are related to $\Lambda$ by hierarchical
dimensionless coupling constants \cite{Foot:2007iy, Foot:2010av}: smaller scales are obtained by
multiplying $\Lambda$ by small dimensionless parameters, while larger scales are obtained by dividing $\Lambda$
by small parameters.  We have demonstrated that mass
hierarchies generated this way can be technically natural, i.e.\ stable against quantum corrections \cite{Foot:2010av}.  
For other related proposals, see \cite{Hempfling:1996ht}.

The cosmological constant (CC) problem has an interesting twist within the scale-invariant framework. 
The bare CC is forbidden by the scale 
invariance of the classical action, and, as has been pointed out in  \cite{Foot:2007iy}, the vanishing of the one-loop 
perturbative contribution requires fine-tuning of the masses of the relevant quantum fields. However, the problem 
with this kind of one-loop fine tuning is that the mass of the dilaton [the (pseudo-) Goldstone boson (PGB) of 
spontaneously broken scale invariance] vanishes also at one-loop.
This is a potential problem, since models with a massless dilaton contradict 
``fifth force" experiments, and thus are phenomenologically unacceptable.

Another potential problem is that the CC is inferred from astronomical observations 
to have a small positive value, and therefore perturbative contribution to CC must be correspondingly balanced with the non-perturbative QCD contribution. However, it turns out that at 1-loop
level the perturbative contribution to CC in scale-invariant models is negative for the desired non-trivial vacuum and hence it cannot 
be cancelled against the known non-perturbative QCD contribution to the CC, because these contributions have 
the same sign  \cite{Foot:2007iy}. Also, since the physical cosmological constant is renormalization-scale 
independent  \cite{Foot:2007wn}, there is no way to ``relax'' it to a small value when measured at large (astronomical) 
scales.\footnote{Here we would like to stress  that the physical CC is understood as  the renormalized vacuum expectation value (VEV) of the energy-momentum tensor that contributes to the right-hand-side of Einstein's equations 
within the semiclassical approach to general relativity.  The renormalization-scale independence 
of the CC is an \emph{exact, non-perturbative} statement, and simply follows from the 
renormalization-scale independence of the effective action, i.e. from the defining property of the 
renormalization group \cite{Foot:2007wn}. Therefore, continuing claims on the possible 
running of the CC \cite{Shapiro:2009dh} are erroneous.} 

It is possible that the problems caused by attempting to incorporate a small positive CC into
scale-invariant theories might be resolved by higher-order terms in the perturbative expansion.
We show in this paper that these terms indeed lift the degeneracy between vanishing CC and vanishing
PGB. Furthermore we show that when higher order terms in the perturbative expansion are included, 
a small positive CC is allowed in the theory without causing any phenomenological problems.
Scale-invariant theories, at least as currently understood, still require a fine tuning of parameters 
in order to accommodate a small positive CC, but they are at least compatible with
a small positive CC.  This is the main point of the paper.

A consequence of incorporating a small positive CC into scale-invariant theories is that the
perturbative contribution must be positive and of order $\Lambda_{QCD}^4$  to approximately cancel
the non-perturbative negative QCD contribution. This then implies that we necessarily live in a metastable
`false' vacuum, since we know the perturbative energy density is zero when
all fields have zero VEVs. We show that the metastable vacuum will not decay on
cosmological time scales for typical scale-invariant models.

\section{Perturbative vacuum energy in scale-invariant models}

Consider a classically scale-invariant theory that 
contains a set of $n$ real scalar fields $S_i$ ($i=1,2,...n$). Some of these scalar fields may form  
multiplets of a local or global symmetry group. The generic classical potential can be written as
\begin{equation}
V_0(S_i)=\lambda_{ijkl}S_iS_jS_kS_l~,
\label{a1}
\end{equation}
where $\lambda_{ijkl}$ are bare coupling constants and summation over the repeated indices is assumed. 
It is convenient to adopt the hyper-spherical parametrization for the scalar fields:
\begin{eqnarray}
S_i(x) &=& r(x)\cos\theta_i(x)\prod_{k=1}^{i-1}\sin\theta_{k}(x)~, \ {\rm for} \ i=1,\ldots,n-1 \nonumber \\
S_n (x) &=& r(x) \prod_{k=1}^{n-1} \sin\theta_k
\label{a2}
\end{eqnarray}
where $r(x)$ is the modulus field. Its nonzero VEV, $\langle r \rangle \neq 0$, 
breaks scale invariance spontaneously resulting in a corresponding (pseudo-)Goldstone boson, the dilaton. 
In the parametrization of (\ref{a2}) the classical potential takes the form
\begin{equation}
V_0(r, \theta_i)=r^4f(\lambda_{ijkl}, \theta_i)~.
\label{a3}
\end{equation} 
Due to the classical scale invariance the modulus field $r(x)$ factors out, 
and the extremum  condition 
$\left.\frac{\partial V_0}{\partial r}\right |_{r=\langle r \rangle,~ \theta_i=\langle \theta_i \rangle}=0$ 
implies, for generic parameter choices, that $\langle r \rangle = 0$.  This in turn implies that
the VEV of the potential, that is, the classical contribution to the CC, vanishes: 
$V_0(\langle r\rangle, \langle \theta_i\rangle )=0$. 

The other way to satisfy the above extremum condition is to have $f(\lambda_{ijkl}, \langle \theta_i \rangle) = 0$,
which would also correspond to a flat direction of $V_0$ that leaves $\langle r \rangle$ undetermined (note that the
value of the classical CC remains at zero).  The dilaton field
would then be massless.  At the classical level, such a flat direction can typically only be achieved by fine tuning
a relation amongst the $\lambda_{ijkl}$.  However, the situation changes at the quantum level.  The coupling constants
then become running parameters depending on the scale $\mu$, and a single relation amongst them is actually
a renormalization condition, not a fine tuning: the condition is obeyed for the specific scale $\mu = \Lambda$,
and this is precisely what dimensional transmutation means. The condition required to achieve a flat direction 
for $V_0$ is used to
define the dimensional transmutation scale.  This is very useful, because along a $V_0$ flat direction quantal
corrections to the tree-level potential will dominate, and VEVs and symmetry breaking patterns can be reliably
computed using perturbation theory, as discussed in the classic paper of Ref.\cite{GildenerWeinberg}.  We shall
use a variation of this technique below.

We now turn to an analysis of the quantized theory.
Quantum corrections lead to an effective potential which can be written in terms of effective, 
renormalization scale $\mu$-dependent couplings and fields,
\begin{eqnarray}
V=A(g_a(\mu), m_{x}(\mu), \theta_i(\mu), \mu) r^{4}(\mu)+B(g_a(\mu), m_{x}(\mu), 
\theta_i(\mu), \mu) r^{4}(\mu)\log\left(\frac{r^2(\mu)}{\mu^2}\right) \nonumber \\
+C(g_a(\mu), m_{x}(\mu), \theta_i(\mu), \mu) r^{4}(\mu)\left[\log\left(\frac{r^2(\mu)}{\mu^2}\right)\right]^2+\ldots~,
\label{a4}
\end{eqnarray}
where $\ldots$ denotes all terms with  higher-power logarithms coming from all possible higher-loop diagrams.  
The parameters $g_a(\mu)$ and $m_x(\mu)$ denote all relevant 
running dimensionless couplings and effective masses, respectively. Equation (4) is the most general form of the 
effective potential for the field $r(x)$ to arbitrary 
order in perturbation theory (see, e.g., the discussion in \cite{Bando:1992wz} and the references therein, and specialize 
to the case of classically scale-invariant theories).  For our purposes it is 
very convenient to fix the renormalization scale as $\mu = \langle r \rangle$. With this choice 
of $\mu$ the higher-power $\log$ terms become irrelevant for our discussion and we 
do not need to display them here.  In addition, since we are primarily interested in the 
VEV of the effective potential (\ref{a4}) we fix the direction of the potential by taking 
$\theta_i = \langle \theta_i \rangle$ in  (\ref{a4}). The extremum condition along the radial direction implies
\begin{eqnarray}
\frac{\partial V}{\partial r} =0 & \Rightarrow & 2A(\mu=\langle r\rangle)  + B(\mu=\langle r\rangle)  = 0 \ .
\label{1}
\end{eqnarray}
If we demand that the perturbative contribution to the CC vanishes, then this
requires that $V_{\min}=0$, that is,
\begin{eqnarray}
V_{\rm min} = 0 & \Rightarrow & A(\mu=\langle r\rangle)  = 0~.
\label{2}
\end{eqnarray}
Note that while $V_{\min}=0$ implies tuning of parameters,
the condition (\ref{1}) is just an extremum condition which simply 
implies that  the scale $\langle r \rangle$ (which is the scale we generically called
$\Lambda$ in the previous section) is defined as the scale $\mu$
where $2A +  B = 0$. Thus, the condition (\ref{1}) trades one dimensionless 
parameter for a dimensional parameter, the phenomenon known as dimensional transmutation.
(This procedure is similar to that of Gildener and Weinberg \cite{GildenerWeinberg} except that we are minimizing the
full effective potential, not just the tree-level potential, in deriving Eq.\ref{1}.  Because large logarithms
are absent, our modified Gildener-Weinberg procedure is consistent with the applicability of a perturbative
approach.)

Evidently, with the above conditions (\ref{1}) and (\ref{2}) the mass of the dilaton $m_{\rm PGB}=
\left.\frac{\partial^ 2V }{\partial r^2}\right |_{r=\mu=\langle r\rangle, \langle\theta_i\rangle}$ 
is determined by at least two-loop level quantum corrections,    
\begin{eqnarray}
m_{\rm PGB}^2 = 8C (\mu = \langle r \rangle) \langle r \rangle^2~.
\label{3}
\end{eqnarray}
Clearly, we must require $C (\mu = \langle r \rangle) > 0$ for the fine tuning of Eq.~(\ref{2}) to be acceptable.

The renormalization group (RG) properties of the effective potential (\ref{a4}) 
give further relations between $A$, $B$ and $C$. 
The potential should not depend on the renormalization scale $\mu$, that is, 
\begin{eqnarray}
\mu\frac{dV}{d\mu}\equiv \left( \mu {\partial \over \partial \mu} + 
\sum_{a} \beta_{a} {\partial \over \partial g_{a}} 
 +\sum_{x}\gamma_{x} m_x {\partial \over \partial m_{x}}- \gamma_r r {\partial \over \partial r}- 
 \sum_{i} \gamma_i \theta_i{\partial \over \partial \theta_i} \right) V = 0~,
\label{4}
\end{eqnarray}
where $\beta_a$ are beta-functions which determine the running of couplings $g_a$, 
while $\gamma_r$, $\gamma_i$ are scalar anomalous 
dimensions and $\gamma_{x}\equiv {\mu \over m_{x}}{\partial m_{x} \over \partial \mu}$ 
are mass anomalous dimensions.  Equations (\ref{1}), (\ref{2}) and (\ref{4}) imply
\begin{eqnarray}
B (\mu = \langle r \rangle) &=& 
\left. {1 \over 2} \mu {d A \over d\mu}\right |_{\mu = \langle r 
\rangle}, \nonumber \\ 
C (\mu = \langle r \rangle) &=& \left. {1 \over 4} \mu {d B \over
d \mu}\right |_{\mu = \langle r \rangle}~.
\label{5}
\end{eqnarray}
The quantities $A$, $B$ and $C$ can in principle be computed in perturbation theory. The leading-order contributions to
$A$, $B$ and $C$ arise at tree $``(0)"$, one loop $``(1)"$, and two loops $``(2)"$, respectively, so that
\begin{eqnarray}
A = A^{(0)} + A^{(1)} + ... \nonumber \\
B = B^{(1)} + B^{(2)} + ... \nonumber \\
C = C^{(2)} + C^{(3)} + ...
\label{6}
\end{eqnarray}
If perturbation theory is valid, then
the conditions $A(\mu = \langle r \rangle) = 0, \ B(\mu = \langle r \rangle) = 0$ and
$C(\mu = \langle r \rangle) > 0$ imply:
\begin{eqnarray}
A^{(0)} (\mu = \langle r \rangle) \approx 0~, \nonumber \\
B^{(1)} (\mu = \langle r \rangle) \approx 0~, \nonumber \\
C^{(2)} (\mu = \langle r \rangle) > 0~.
\label{7}
\end{eqnarray}
Again, the first condition can be used simply to define approximately the scale $\mu = \langle r \rangle$,
and results in the elimination of one of the tree-level parameters in the potential.
The quantity $B^{(1)}$ is in general
\begin{eqnarray}
B^{(1)} (\mu = \langle r \rangle) =\left. {1 \over 64 \pi^2 
\langle r \rangle^4} [3{\rm Tr} m_V^4 + {\rm Tr} m_S^4 - 4{\rm Tr}
m_F^4]\right |_{\mu = \langle r \rangle}~,
\label{8}
\end{eqnarray}
where the subscripts $V$, $S$ and $F$ denote contributions of massive vector bosons, 
scalars and Dirac fermions, respectively. The quantity $C^{(2)}$ then is given by
\begin{eqnarray}
C^{(2)} (\mu = \langle r \rangle) =\left. {1 \over 64 \pi^2 \langle r \rangle^4} \left[ 3 {\rm Tr} m_V^4 \gamma_V + {\rm Tr}
m_S^4\gamma_S - 4 {\rm Tr} m_F^4 \gamma_F \right]\right |_{\mu = \langle r \rangle}~, 
\label{9}
\end{eqnarray}
where $\gamma_x=\frac{\partial \ln m_x}{\partial \ln \mu}$ $(x=V,S,F)$, and we have used (\ref{5}), (\ref{8}) and (\ref{7}) [note that although $B^{(1)}$ in (\ref{8}) is evaluated at $\mu=\langle r\rangle$, the formulae in (\ref{8}) holds for an arbitrary $\mu$]. \emph{ A priori}, $C^{(2)}$ in (\ref{9}) is not positive, 
thus the condition  $C^{(2)}>0$ puts a restriction on the particle spectrum of the theory. 

Since the CC is a relevant observable only in the presence of gravity, we would like to briefly
mention also how gravity can be incorporated within this kind of framework. The simplest
way is to assume that the Planck mass is spontaneously generated (see, e.g., \cite{Zee:1978wi}) 
through the scale-invariant non-minimal couplings 
\begin{equation}
\sqrt{-g}\xi_{ij}S_iS_j R, 
\label{9a}
\end{equation}
where $R$ is the Ricci scalar and $\xi_{ij}$ are dimensionless coupling constants.  We shall use this idea shortly.

Let us now apply this formalism to a particular model. We consider the model with a Higgs doublet and
two real singlet scalar fields \cite{Foot:2010av}, which was identified as the simplest perturbative model capable of 
explaining the various scales: electroweak, see-saw and Planck scales. In that model,
the two real scalars gain large VEVs which generate the Planck scale and see-saw mass scale,
and a tiny coupling to the Higgs doublet generates the electroweak scale.
Consider the part of the tree-level potential involving the two real scalar fields,
\begin{eqnarray}
V_0 (S_1, S_2) = {\lambda_1 \over 4} S_1^4 + {\lambda_2 \over 4} S_2^4 + {\lambda_3 \over 2} S_1^2
S_2^2~.
\end{eqnarray}
We parametrize the fields via
\begin{eqnarray}
S_1 = r\cos \theta, \ S_2 = r\sin\theta~,
\end{eqnarray}
and we choose the $\lambda_3 < 0$ parameter space.
In this case, $V_0 (r) = A^{(0)} r^4$ and $A^{0} (\mu = \langle r \rangle) = 0$ implies
\begin{eqnarray}
\langle S_1 \rangle = \langle r \rangle \left( { 1 \over 1 + \epsilon}\right)^{1/2} \equiv v,
\ \langle S_2 \rangle = v\epsilon^{1/2},
\end{eqnarray}
where
\begin{eqnarray}
\langle \theta \rangle=\omega~,~~\tan^2 \omega  \equiv \epsilon = \sqrt{{\lambda_1 (\mu) 
\over \lambda_2
(\mu)}}~,
\label{aa}
\end{eqnarray}
with
\begin{eqnarray}
\lambda_3 (\mu) + \sqrt{ \lambda_1 (\mu) \lambda_2 (\mu)} = 0
\end{eqnarray}
and $\mu = \langle r \rangle$.

In this model, $\langle S_1 \rangle$ sets the Planck scale while $\langle S_2 \rangle$ sets the see-saw scale
via the standard Lagrangian terms,
\begin{eqnarray}
{\cal L} = \sqrt{-g} \xi S_1^2 R + \sqrt{-g}\sum_{i=1}^3\lambda^i_M \bar \nu_{i R} (\nu_{i R})^c S_2+{\rm h.c.}~
\end{eqnarray}
Clearly, $m_{iF} = 2\lambda^i_M \langle S_2\rangle$. The two mass eigenstate scalars are denoted by $S=\cos\omega S_2 - \sin\omega S_1, 
\ s = \sin\omega S_2 + \cos\omega S_1$,
with $m_S^2 = 2\lambda_2 \langle S_2 \rangle^2/\cos^2 \omega$ (while $s$ is the PGB which
gains mass at two loop level, so that $m_s^2 \ll m_S^2$).
Thus, $B^{(1)} \approx 0$ implies
$m_S^4 \approx 2\sum_{i=1}^3 m_{iF}^4$.
Evaluating the anomalous mass dimensions for the scalar $S$ and for $N$ degenerate right-handed
neutrinos in the relevant parameter regime where $\cos^2 \omega \simeq 1$, we find:
\begin{eqnarray}
\gamma_S &=& {3\lambda_2 \over 4\pi^2} 
- {3N \lambda^2_M \over 2\pi^2} \left( {m_{F}^2 \over m_{S}^2} - {1 \over 6}\right)
\nonumber \\
\gamma_F &=& {3\lambda_M^2 \over 4\pi^2}\ .
\end{eqnarray}
Using $\lambda_M^2 = m_{F}^2/(4\langle S_2 \rangle^2)$ and defining $y \equiv 2N m^4_{F}/m_S^4$
(note that $B^{(1)} = 0 \Rightarrow y \simeq 1$), we find
\begin{eqnarray}
C^{(2)} = {3 \lambda_1 \lambda_2^2 \over 128 \pi^4} \left[ 
2 - y + {\sqrt{2N y} \over 6}
- y \sqrt{ {y \over 2N}} \right]\ .
\end{eqnarray}
Evidently $C^{(2)} > 0$ independently of $N$. A similar conclusion can be reached in 
the more general case of non-degenerate right-handed neutrinos   
and hence the model is consistent with the inferred small CC.   
The PGB mass can then be
estimated from Eq.(\ref{3}). Clearly there is a large range of parameters where the PGB mass
is greater than the limits suggested by ``fifth force" experiments (typical limits
from such experiments are for PGB masses greater than of order $10^{-2}$ eV).

In this model, the VEV of the standard Higgs doublet arises from the tiny couplings
\begin{eqnarray}
V = \lambda_x^i \phi^{\dagger} \phi S^2_i \ .
\end{eqnarray}
Reanalyzing the model incorporating such small couplings will not significantly change any of the above
considerations. Note also, that perturbative contributions from the Standard Model fields to the CC are negligible compared to the contributions of the heavy hidden sector fields.  They result in a slight modification of our hidden sector mass relations by contributions of the order of $\sim {\cal O}(m_{\rm t})$ at most.

\section{Living in a metastable vacuum}

Besides the perturbative contribution to the vacuum energy density discussed in the previous section, 
there is a non-perturbative contribution from the confining phase of QCD.  
It is dominated by the non-perturbative condensate expectation value of the gluonic operator \cite{Shifman:1978bx},
\begin{equation}
\langle {\rm Tr}G_{\mu\nu}G^{\mu\nu}\rangle=-\Lambda_{\rm QCD}^4~,
\label{b1}
\end{equation}
 where $\Lambda_{\rm QCD}\sim 300$ MeV. This contributes to the total vacuum energy density with a negative sign, 
 \begin{equation}
 E_{\rm vac}=V_{\rm min}-\Lambda_{\rm QCD}^4~.
 \label{b2}
 \end{equation}
Thus, taking that $E_{\rm vac} \approx \Lambda^4\sim 10^{-120} M_{\rm P}^4$ 
[$M_{\rm P}\approx 1.2\cdot 10^{19}$ GeV is the Planck mass], as can be inferred from astrophysical observations, 
we find that the perturbative contribution must be positive 
$V_{\rm min}=A\langle r \rangle^4\vert_{\mu=\langle r\rangle}\approx\Lambda_{\rm QCD}^4$. 
This slightly modifies the particle spectrum since the extremum 
condition (5) implies that $B(\mu=\langle r\rangle)\approx -2\Lambda_{\rm QCD}^4/\langle r\rangle^ 4\neq 0$. 
Such a modification does not significantly affect the PGB mass (7), 
which now reads: $m_{\rm PGB}^ 2= 8\left[ B(\mu=\langle  r\rangle)+C(\mu=\langle r\rangle)\right] \langle r\rangle^2
\approx - 16 \frac{\Lambda_{QCD}^4}{\langle r\rangle^ 2}+ 8C(\mu=\langle r\rangle)\langle r\rangle^2 
\simeq 8C(\mu=\langle r\rangle)\langle r\rangle^2$, 
since $\langle r\rangle/\Lambda_{\rm QCD}\gg 100$ in realistic models. 
Note, however the important role of the 2-loop contribution $C$, in the absence of which one might conclude 
that the small positive CC is incompatible with the desired non-trivial vacuum (the dilaton is tachyonic in this case) 
and only negative CC can be accommodated.   

Although a small value for $V_{\rm min}$ does not affect significantly our analysis in the 
previous section, we are driven now to the conclusion that we live in a false vacuum (a local minimum of the effective potential), with the true vacuum having energy $E_{\rm true~vac}=\left.V_{\rm min}\right |_{r=0}-\Lambda_{\rm QCD}^4
\approx -\Lambda_{\rm QCD}^4$. The false vacuum obviously must have a long enough lifetime, 
and this puts constraints on possible QCD-type non-perturbative contributions in our theory. 
Indeed, if the decay rate per unit volume of a false vacuum is larger than 
$\Gamma_{\rm cr}= H^4_{\rm today}\sim {{E_{\rm vac}^2 }\over M_P^4}\sim 10^{-240}M_{\rm P}^4$, 
the bubbles of ``true" vacuum will collide and eventually fill the 
entire visible  Universe. For a slower decay rate, the expansion of the Universe dominates the 
proliferation of the  true vacuum, and the visible Universe remains in the metastable vacuum state.      

The decay rate of a false vacuum per unit volume  in the semi-classical approximation is given by \cite{Coleman:1977py} 
\begin{equation}
\Gamma \sim \frac{1}{R^4}{\rm e}^{-S_{\rm E}}~,
\label{b3}
\end{equation}
with $R$ being the ``4D Euclidean size" of the bubble that maximizes the rate (\ref{b3}) and $S_{\rm E}$ is the Euclidean action 
evaluated on the classical bounce solution for the modulus field $r(x)$.  
In the thin-wall approximation, which is valid in our case, Coleman \cite{Coleman:1977py} found that
\begin{eqnarray}
R &=& {3\sigma \over \epsilon}~, \nonumber \\
S_{\rm E} &=& {27\pi^2 \sigma^4 \over 2\epsilon^3}~,
\end{eqnarray}
where $\epsilon \approx \Lambda^4_{QCD}$ is the false-true vacuum energy difference and 
$\sigma$ is
the tension of a domain wall separating true and false vacua \cite{Coleman:1977py}:
\begin{equation}
\sigma=\int_0^{\langle r \rangle} dr  \sqrt{2 V}=\frac{1}{9}m_{\rm PGB}\langle r\rangle^2~. 
\label{b6}
\end{equation}
Thus we find that
\begin{eqnarray}
S_{\rm E}\approx \frac{\pi^2}{486}\frac{m_{\rm PGB}^4\langle r \rangle^8}{\Lambda_{\rm QCD}^{12}}~.
\end{eqnarray}
The condition that the electroweak vacuum is metastable, $\Gamma \stackrel{<}{\sim} \Gamma_c$, then implies that
\begin{eqnarray}
\left(\frac{m_{\rm PGB}}{\rm GeV}\right)^4 \left(\frac{0.3~{\rm GeV}}{\Lambda_{QCD}}\right)^{12}
\stackrel{>}{\sim} - 10^{-153}
\log
\left(
\left[\frac{m_{\rm PGB}}{\rm GeV}\right ]^4 \left[\frac{0.3~{\rm GeV}}{\Lambda_{QCD}}\right]^{16} 
\right)~.
 \label{b9}
\end{eqnarray}
This condition is satisfied for all reasonable choices of the PGB mass.
However, models with new extra QCD-type 
confinement scales are constrained. In fact, models with  $\Lambda_{\rm QCD'}\stackrel{>}{\sim} 10^{13} 
\left(\frac{m_{\rm PGB}}{\rm GeV}\right)^{1/3}$ GeV are excluded.

\section{Conclusion}

In this paper we have discussed the cosmological constant problem within realistic scale-invariant models. 
The vacuum energy, and thus the cosmological constant, in scale-invariant theories vanishes at the classical 
level but is induced radiatively alongside the anomalous breaking of scale invariance. 
A vanishing perturbative contribution to the cosmological constant requires fine adjustment of the 
masses of particles. This, in turn, implies that  the PGB dilaton acquires its mass at two-loop level, 
and hence $C^{(2)}$ (\ref{9}) must be positive in order for the scale invariant breaking vacuum to be the 
minimum of the effective potential (\ref{a4}). This  holds true in the simplest scale-invariant 
model of Ref.~\cite{Foot:2010av} where the hierarchies between the electroweak, 
see-saw and Planck scales are incorporated in a technically natural way. 
Taking into account  the nonperturbative QCD contribution to the cosmological constant, 
we found that the desired vacuum is actually a local minimum, but not a global one. 
Nevertheless, this vacuum turns out to be very long-lived for reasonable values of the 
parameters of the theory. Therefore, a small positive cosmological constant 
can be consistently incorporated within radiatively-broken scale-invariant models.

\subsection*{Acknowledgements}

This work was supported in part by the Australian Research Council.


\begin{thebibliography}{99}
  
\bibitem{Foot:2007as}
  R.~Foot, A.~Kobakhidze and R.~R.~Volkas,
  Phys.\ Lett.\  B {\bf 655}, 156 (2007)
  [arXiv:0704.1165 [hep-ph]].

\bibitem{Foot:2007ay}
  R.~Foot, A.~Kobakhidze, K.~L.~McDonald and R.~R.~Volkas,
  Phys.\ Rev.\  D {\bf 76}, 075014 (2007)
  [arXiv:0706.1829 [hep-ph]].

\bibitem{Foot:2007iy}
  R.~Foot, A.~Kobakhidze, K.~L.~McDonald and R.~R.~Volkas,
  Phys.\ Rev.\  D {\bf 77}, 035006 (2008)
  [arXiv:0709.2750 [hep-ph]].

\bibitem{Foot:2010av}
  R.~Foot, A.~Kobakhidze and R.~R.~Volkas,
  Phys.\ Rev.\  D {\bf 82}, 035005 (2010)
  [arXiv:1006.0131 [hep-ph]].


\bibitem{Coleman:1973jx}
  S.~R.~Coleman and E.~J.~Weinberg,
  Phys.\ Rev.\  D {\bf 7}, 1888 (1973).


  \bibitem{Hempfling:1996ht}
  R.~Hempfling,
  Phys.\ Lett.\  B {\bf 379}, 153 (1996)
  [arXiv:hep-ph/9604278]; 
  W.~F.~Chang, J.~N.~Ng and J.~M.~S.~Wu,
  Phys.\ Rev.\  D {\bf 75}, 115016 (2007)
  [arXiv:hep-ph/0701254];
  T.~Hambye and M.~H.~G.~Tytgat,
  Phys.\ Lett.\  B {\bf 659}, 651 (2008)
  [arXiv:0707.0633 [hep-ph]]; 
  S.~Iso, N.~Okada and Y.~Orikasa,
  Phys.\ Lett.\  B {\bf 676}, 81 (2009)
  [arXiv:0902.4050 [hep-ph]];
  M.~Holthausen, M.~Lindner and M.~A.~Schmidt,
  arXiv:0911.0710 [hep-ph];
  L.~Alexander-Nunneley and A.~Pilaftsis,
  JHEP {\bf 1009}, 021 (2010)
  [arXiv:1006.5916 [hep-ph]]; 
  A.~Latosinski, K.~A.~Meissner and H.~Nicolai,
  arXiv:1010.5417 [hep-ph].

\bibitem{Foot:2007wn}
  R.~Foot, A.~Kobakhidze, K.~L.~McDonald and R.~R.~Volkas,
  Phys.\ Lett.\  B {\bf 664}, 199 (2008)
  [arXiv:0712.3040 [hep-th]].

\bibitem{Shapiro:2009dh}
  I.~L.~Shapiro and J.~Sola,
  Phys.\ Lett.\  B {\bf 682}, 105 (2009)
  [arXiv:0910.4925 [hep-th]]; 
  B.~F.~L.~Ward,
  Mod.\ Phys.\ Lett.\  A {\bf 25}, 607 (2010)
  [arXiv:0908.1764 [hep-ph]];
  S.~Domazet and H.~Stefancic,
  arXiv:1010.3585 [gr-qc].

\bibitem{GildenerWeinberg}
E.~Gildener and S.~Weinberg, Phys.\ Rev.\ {\bf D13}, 3333 (1976).


\bibitem{Bando:1992wz}
  M.~Bando, T.~Kugo, N.~Maekawa and H.~Nakano,
  Phys.\ Lett.\  B {\bf 301}, 83 (1993)
  [arXiv:hep-ph/9210228].

\bibitem{Zee:1978wi}
  A.~Zee,
  Phys.\ Rev.\ Lett.\  {\bf 42 } (1979)  417.


\bibitem{Shifman:1978bx}
  M.~A.~Shifman, A.~I.~Vainshtein and V.~I.~Zakharov,
  Nucl.\ Phys.\  B {\bf 147}, 385 (1979).


\bibitem{Coleman:1977py}
  S.~R.~Coleman,
  Phys.\ Rev.\  D {\bf 15}, 2929 (1977)
  [Erratum-ibid.\  D {\bf 16}, 1248 (1977)]. See also the earlier works:  
  I.~Y.~Kobzarev, L.~B.~Okun and M.~B.~Voloshin,
  Sov.\ J.\ Nucl.\ Phys.\  {\bf 20}, 644 (1975)
  [Yad.\ Fiz.\  {\bf 20}, 1229 (1974)].
  P.~H.~Frampton,
  Phys.\ Rev.\ Lett.\  {\bf 37}, 1378 (1976)
  [Erratum-ibid.\  {\bf 37}, 1716 (1976)].


\end{thebibliography}
\end{document}